\documentclass[reprint,amsmath,amssymb,aps,noeprint,floatfix, nolinenumbers,longbibliography]{revtex4-2}

\usepackage[english]{babel}
\usepackage[utf8]{inputenc}
\usepackage[colorinlistoftodos, color=green!40, prependcaption]{todonotes}
\usepackage{float}
\usepackage{graphicx}
\usepackage{yhmath}
\usepackage{subfigure}
\usepackage{mathdots}
\usepackage{MnSymbol}
\usepackage{dsfont}
\usepackage{soul}
\usepackage[normalem]{ulem}
\usepackage{hyperref}

\hypersetup{
    unicode=true, 
    plainpages=false,
    colorlinks=true,
    linkcolor=blue,
    citecolor=blue,
    filecolor=blue,
    urlcolor=blue
}

\usepackage{mathtools}

\DeclarePairedDelimiterX\braket[2]{\langle}{\rangle}{#1\,\delimsize\vert\,\mathopen{}#2}

\newcommand{\md}{\mathrm{d}}

\newcommand{\sgn}{\mathrm{sgn}}

\begin{document}

\title{A Fully Analog Pipeline for Portfolio Optimization}

\author{James S. Cummins}
\email[correspondence address: ]{jsc95@cam.ac.uk}
\author{Natalia G. Berloff}
\affiliation{Department of Applied Mathematics and Theoretical Physics, University of Cambridge, Wilberforce Road, Cambridge CB3 0WA, United Kingdom}

\date{\today}

\begin{abstract}
Portfolio optimization is a ubiquitous problem in financial mathematics that relies on accurate estimates of covariance matrices for asset returns. However, estimates of pairwise covariance could be better and calculating time-sensitive optimal portfolios is energy-intensive for digital computers. We present an energy-efficient, fast, and fully analog pipeline for solving portfolio optimization problems that overcomes these limitations. The analog paradigm leverages the fundamental principles of physics to recover accurate optimal portfolios in a two-step process. Firstly, we utilize equilibrium propagation, an analog alternative to backpropagation, to train linear autoencoder neural networks to calculate low-rank covariance matrices. Then, analog continuous Hopfield networks output the minimum variance portfolio for a given desired expected return. The entire efficient frontier may then be recovered, and an optimal portfolio selected based on risk appetite.
\end{abstract}

\maketitle

\section{Introduction}

Portfolio optimization involves creating an investment portfolio that balances risk and return. The objective is to allocate assets optimally to maximize expected returns while minimizing risk. Naturally, this problem is of great interest to financial organizations and is pivotal in risk management. However, the problem, formulated by Markowitz's mean-variance model \cite{markowitz1952portfolio}, must be solved in practice. Namely, it is well known that estimates of pairwise covariance between assets are notoriously poor \cite{ledoit2022power}. A large financial company may have hundreds of thousands of assets $n$ covering equities, bonds, derivatives, and more, but with only a small sample of observations over the desired timescale. The samples tend to include significant amounts of noise, distorting the underlying relationships between the assets. The symmetric covariance matrix has $n (n + 1) / 2$ total unique terms: $n (n - 1 ) / 2$ pairwise correlations and $n$ variances. Hence, the number of unique terms behaves as $\mathcal{O} (n^{2})$, which leads to significant potential for an ill-conditioned covariance matrix \cite{tabeart2020improving}. To overcome this issue, factor models were introduced that vastly reduce the dimensionality, and thus the number of numerical estimates required \cite{bai2002determining}. Factor methods produce low-rank covariance matrices that retain only the largest eigenvalues and discard small eigenvalues associated with noise. Despite this development, the computation of optimal portfolios remains energy-intensive as the efficient frontier is mapped out in $n$ dimensions. In high-frequency trading, this becomes a time-sensitive computation as assets are purchased and sold on microsecond timescales, and portfolios must be regularly rebalanced not to exceed risk appetites. Much attention has been focused on portfolio optimization in the high-frequency domain \cite{liu2009portfolio, goumatianos2013stock, ziegelmann2015selection}, including the use of evolutionary algorithms to update efficient frontiers \cite{filipiak2017dynamic}. By using such fundamental principles as minimizing entropy, energy, and dissipation \cite{vadlamani2020physics}, or, perhaps,   incorporating quantum phenomena like superposition and entanglement \cite{farhi}, we can advance and surpass the classical computations of these problems. At the forefront of this drive to alternate architectures is the integration of analog, physics-based algorithms and hardware, which involve translating complex optimization problems into universal spin Hamiltonians \cite{lucas2014ising, berloff2017realizing, kalinin2020polaritonic}. Indeed, the mean-variance portfolio optimization framework can be encoded into a Hamiltonian's coupling strengths with the physical system recovering the Hamiltonian's ground state, which corresponds to the optimal portfolio solution \cite{beasley2013portfolio, wang2024efficient}. Efficient mapping from the original problem description to spin Hamiltonian enables the problem to remain manageable despite increasing complexity \cite{Cubitt_universality}.

In Section \ref{Mean-Variance Optimization}, we introduce the mean-variance optimization framework for calculating optimal portfolios. Then, in Section \ref{Continuous Hopfield Network}, we show that analog continuous Hopfield networks can solve portfolio optimization problems by evolving to the minimum of an energy function that encodes the problem parameters. In Section \ref{Low-Rank Approximation}, we address the issues of estimating pairwise covariance by introducing the low-rank approximation that relies on a low-dimensional latent variable representation. In Section \ref{Linear Autoencoders}, we show that calculating such a representation can be done using linear autoencoder neural networks, and in Section \ref{Equilibrium Propagation} how these networks can be trained on analog hardware using equilibrium propagation. Section \ref{Results} brings everything together, starting with  raw data observations and working through the entire analog pipeline.

\section{Mean-Variance Optimization}
\label{Mean-Variance Optimization}

\begin{figure}[t]
\centering
     \includegraphics[width=\columnwidth]{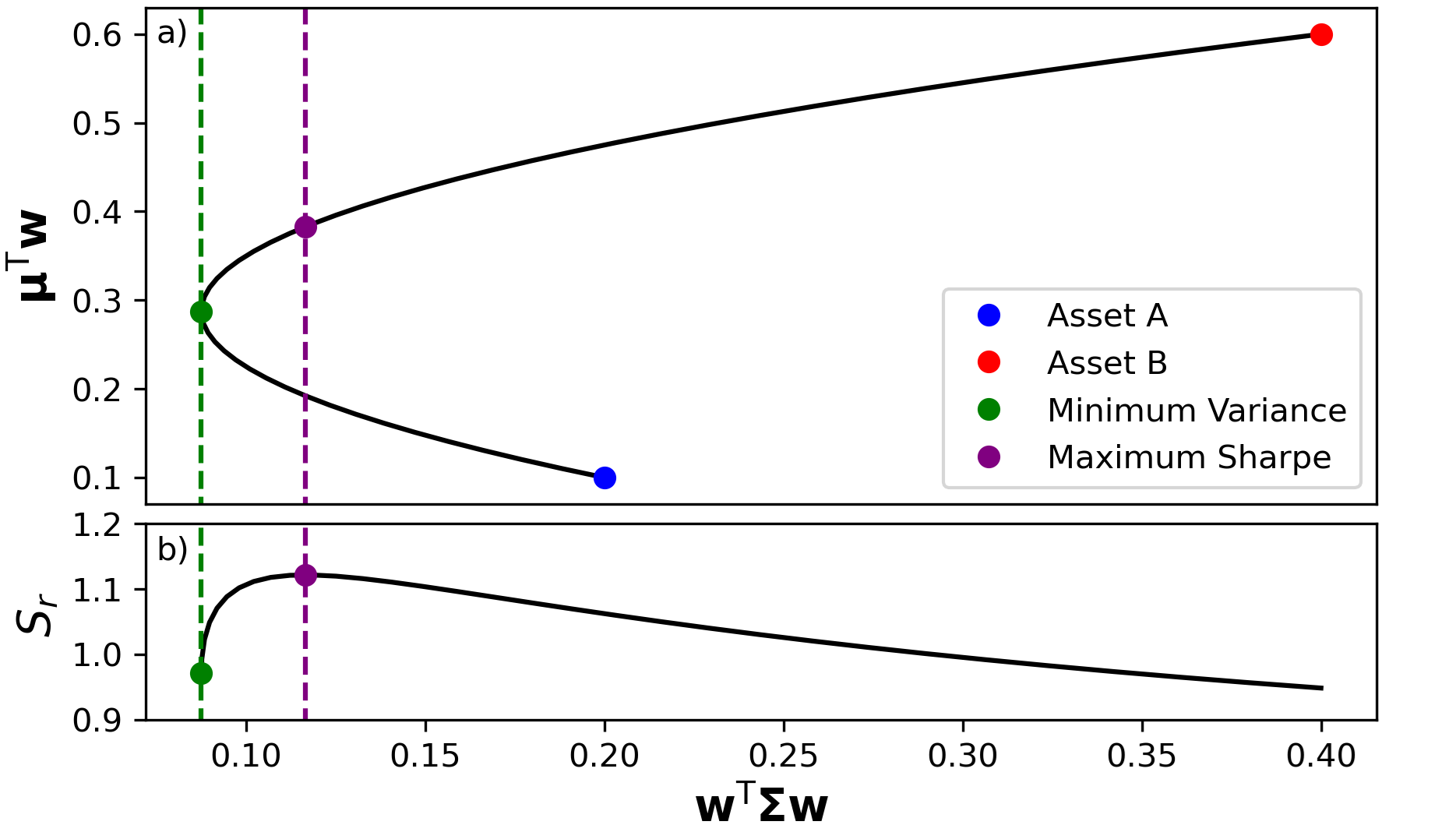}
     \caption{(a) The hyperbola in variance-return space for a portfolio of $n = 2$ assets $A$ and $B$. The positively sloped portion of this hyperbola is the efficient frontier. The expected returns are $\mu_{A} = 0.1$ and $\mu_{B} = 0.6$. The (co)variances are $\sigma_{AA} = 0.2$, $\sigma_{BB} = 0.4$, and $\sigma_{AB} = \sigma_{BA} = -0.1$. The blue circle represents the portfolio consisting only of asset $A$, and the corresponding investment weights are $\mathbf{w} = [1, 0]^{\rm T}$. Likewise, the red circle is the portfolio consisting only of asset $B$. The minimum variance portfolio, shown as a green circle, is the combination of weights $\mathbf{w}$ that minimizes the total variance $\mathbf{w}^{\rm T} \mathbf{\Sigma} \mathbf{w}$. The purple circle is the portfolio that maximizes the Sharpe ratio $S_{r}$. The Sharpe ratio is a measure of risk-adjusted return and is defined as $S_{r} = \boldsymbol{\mu}^{\rm T} \mathbf{w} / \sqrt{\mathbf{w}^{T} \mathbf{\Sigma} \mathbf{w}}$. (b) The Sharpe ratio $S_{r}$ for each portfolio in the efficient frontier. We now see that the purple circle is indeed the portfolio that maximizes the Sharpe ratio.}
    \label{Frontier Plot}
\end{figure}

We define $\mu_i$ as the expected return of asset $i$, and $[ \mathbf{\Sigma} ]_{ij} = \sigma_{ij} = {\rm Cov} (i, j)$ as the covariance between assets $i$ and $j$. The decision variables are $w_i$, the proportion of the total investment in asset $i$. For a universe of securities with $n$ assets, the Markowitz mean-variance portfolio optimization problem is
\begin{equation} \label{Portfolio Optimization}
\begin{aligned}
\min_{\mathbf{w}} \quad &  \mathbf{w}^{\rm T} \mathbf{\Sigma} \mathbf{w} \\
\textrm{s.t.} \quad & \boldsymbol{\mu}^{\rm T} \mathbf{w} = R, \\
& \mathbf{1}^{\rm T} \mathbf{w} = 1, \\
& 0 \leq w_i \leq 1,
\end{aligned}
\end{equation}
for $i = 1, \dots, n$, and the condition $w_i \geq 0$ prohibits shorting \cite{beasley2013portfolio}. The variance $\mathbf{w}^{\rm T} \mathbf{\Sigma} \mathbf{w}$ quantifies the portfolio risk for positive semidefinite matrix $\mathbf{\Sigma}$, while $R$ is the desired expected return of the portfolio. $\boldsymbol{\mu}$ and $\mathbf{\Sigma}$ are not known a priori and must be estimated from historical data. The efficient frontier is calculated by solving (\ref{Portfolio Optimization}) for various $R$. The efficient frontier is the set of portfolios that minimize the risk for a given $R$. We illustrate a frontier in Fig.~(\ref{Frontier Plot}) for a toy model with $n = 2$ assets. It was recently suggested that portfolio optimization problems could be solved on analog spatial-photonic Ising machines for equal-weighted portfolios, that is, $w_i \in \{ 0, 1/q \}$ for $q$ selected assets \cite{wang2024efficient}. We go beyond this constraint by utilizing analog Hopfield networks and consider the quadratic continuous optimization problem (\ref{Portfolio Optimization}). In Section \ref{Continuous Hopfield Network}, we aim to recover the optimal asset weights $\mathbf{w}$, given known parameters $\boldsymbol{\mu}$ and $\mathbf{\Sigma}$.

\section{Continuous Hopfield Network}
\label{Continuous Hopfield Network}

\begin{figure}[ht]
\centering
     \includegraphics[width=\columnwidth]{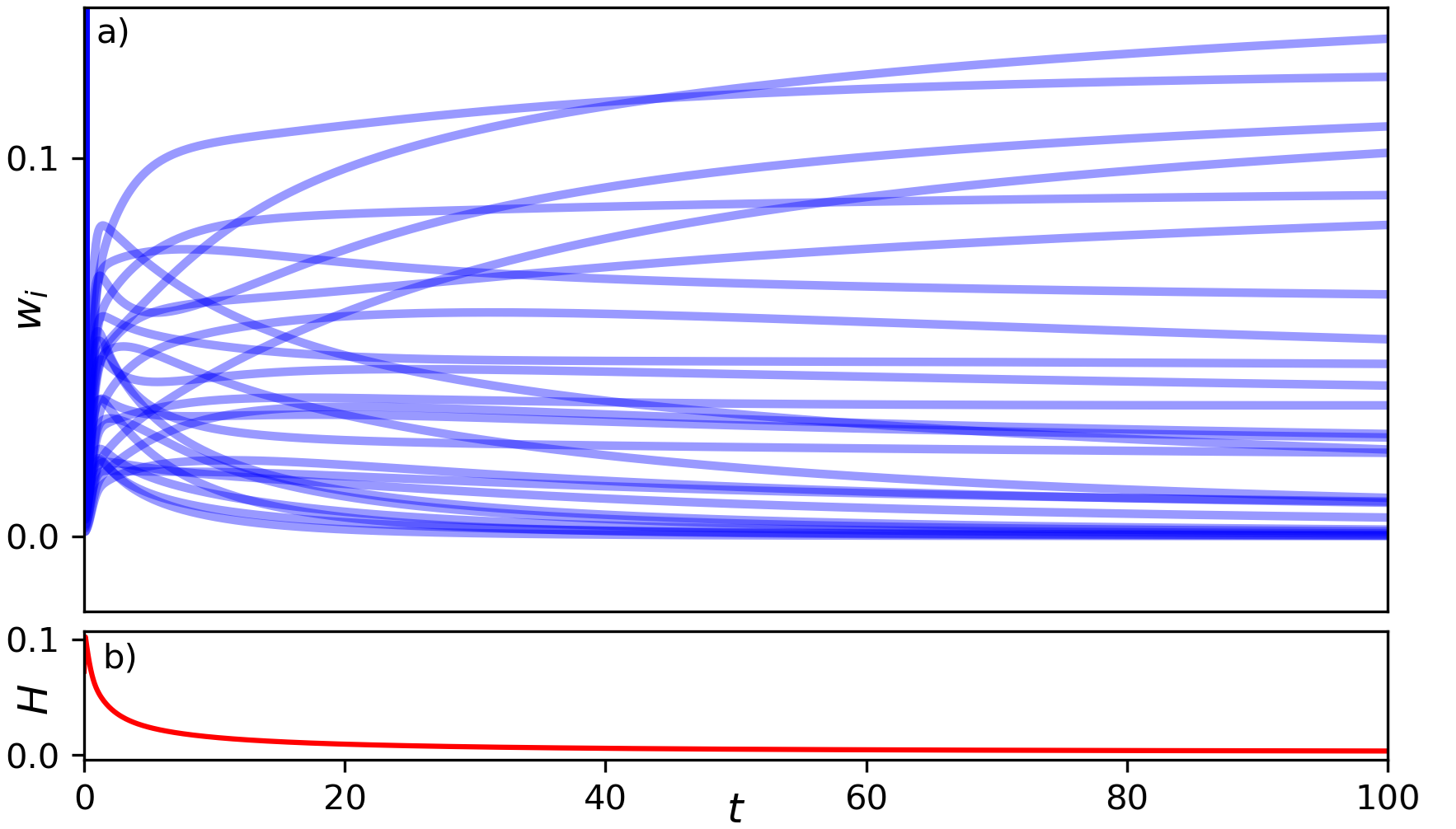}
     \caption{(a) Hopfield network dynamics for a portfolio of $n = 25$ assets with $R = \lambda_{1} = \lambda_{2} = 1$. The dynamical system evolves according to Eq.~(\ref{Hopfield}), which in turn minimizes Eq.~(\ref{Equivalent}). Each line represents one asset weight $w_{i}$. (b) The value of expression (\ref{Equivalent}) during the network dynamics. Covariance matrix $\mathbf{\Sigma}$ and expected return vector $\boldsymbol{\mu}$ are calculated from sampling $N = 10$ observations of returns $\mathbf{x}$ from IID random normal variables $x_{j} \sim N(1, 1)$, where $j = 1, 2, \ldots, N$. The low number of observations $N$ results in a noisy positive semidefinite covariance matrix $\mathbf{\Sigma}$ whose pairwise entries $\sigma_{ij}$ are nonzero. The externally controlled annealed parameter has form $p(t) = p_{0} ( 1 - t / T )$, where $T$ is the total annealing period. Here, $p_{0} = 0.01$ and $T = 100$.}
    \label{Hopfield Plot}
\end{figure}

A continuous Hopfield network is a type of Hopfield neural network which has continuous states and dynamics \cite{hopfield1984neurons}. It is an analog computational network for solving optimization problems. For a network of size $n$, the $i$-th network element at time $t$ is described by a real input $x_{i} (t)$, and the network dynamics are governed by
\begin{equation} \label{Hopfield}
    \frac{\md x_{i}}{\md t} = - p(t) x_{i} + \sum_{j = 1}^{n} J_{ij} v_{j} + m_{i},
\end{equation}
where $v_{i} = g (x_{i})$ is a nonlinear activation function, $p(t)$ is an annealing parameter, $m_{i}$ are the offset biases, and $J_{ij} = [ \mathbf{J} ]_{ij}$ are elements of the symmetric coupling matrix $\mathbf{J}$. Should $g(x)$ be a non-decreasing function, then the steady states of the continuous Hopfield network (\ref{Hopfield}) are the minima of the Lyapunov function
\begin{equation} \label{HT_Energy}
    E = p(t) \sum_{i = 1}^{n} \int_{0}^{v_i} g^{-1} (x) \md x - \frac{1}{2} \sum_{i, j = 1}^{n} J_{ij} v_{i} v_{j} - \sum_{i = 1}^{n} m_{i} v_{i}.
\end{equation}
We choose the functional form of $g(x)$, such that when $p(t) \to 0$, the minima of $E$ occur for $v_{i} \in [0, 1]$ and correspond to the minima of $- \mathbf{v}^{\rm T} \mathbf{J} \mathbf{v}$. Therefore, by setting $\mathbf{J} = - \mathbf{\Sigma}$, we can minimize the variance $\mathbf{w}^{\rm T} \mathbf{\Sigma} \mathbf{w}$ of problem (\ref{Portfolio Optimization}). To satisfy the constraints in problem (\ref{Portfolio Optimization}) we introduce Lagrange multiplier-like scalars $\lambda_{1}$, $\lambda_{2}$ and seek to minimize the expression
\begin{equation}
   H = \mathbf{w}^{\rm T} \mathbf{\Sigma} \mathbf{w} + \lambda_{1} (\boldsymbol{\mu}^{\rm T} \mathbf{w} - R)^2 + \lambda_{2} (\mathbf{1}^{\rm T} \mathbf{w} - 1)^{2}.
\end{equation}
Therefore, after discarding constants, we seek to minimize
\begin{equation} \label{Equivalent}
    H = - \frac{1}{2} \mathbf{w}^{\rm T} \mathbf{J} \mathbf{w} - \mathbf{m}^{\rm T} \mathbf{w},
\end{equation}
where $\mathbf{J} = - 2 \mathbf{\Sigma} - 2 \lambda_{1} \boldsymbol{\mu} \boldsymbol{\mu}^{\rm T} - 2 \lambda_{2} \mathbf{1} \mathbf{1}^{\rm T}$, and $\mathbf{m} = 2 R \lambda_{1} \boldsymbol{\mu} + 2 \lambda_{2} \mathbf{1}$. Equation (\ref{Equivalent}) can be directly encoded into the Hopfield network (\ref{Hopfield}), and if required, $\mathbf{m}$ can be absorbed into $\mathbf{J}$ by introducing an additional auxiliary spin. The non-decreasing monotonic function $g ( x )$ is chosen to be the logistic function $g ( x ) = 1 / [ 1 - \exp ( - x) ]$ to limit possible values of $v_{i}$ such that $0 \leq v_{i} \leq 1$.
We illustrate the Hopfield network dynamics in Fig.~(\ref{Hopfield Plot}) for a randomly generated covariance matrix $\mathbf{\Sigma}$ and expected return vector $\boldsymbol{\mu}$. The energy minimization properties of Hopfield networks make them particularly suitable for solving combinatorial optimization problems. Further extensions have been proposed to increase convergence to optimal states in challenging optimization problems. For example, the first-order Eq.~(\ref{Hopfield}) can be momentum-enhanced and replaced with a second-order equation leading to Microsoft's analog iterative machine \cite{kalinin2023analog} or Toshiba's bifurcation machine \cite{goto2016bifurcation}.

\section{Low-Rank Approximation}
\label{Low-Rank Approximation}

We now focus on calculating a low-rank approximation of the covariance matrix, which will be used in (\ref{Portfolio Optimization}). If $\mathbf{x}_i \in \mathbb{R}^{n}$ are the $i$-th sample of asset returns over $N$ total samples, and we assume that $\mathbb{E} [ \mathbf{x} ] = \mathbf{0}$, then the sample covariance matrix is
\begin{equation} \label{Sample Covariance}
    \mathbf{S} = \frac{1}{N} \sum_{i = 1}^{N} \mathbf{x}_i \mathbf{x}_{i}^{\rm T}.
\end{equation}
When the number of samples $N$ is of the same magnitude as $n$, then the sample covariance matrix usually suffers a large estimation error \cite{ledoit2022power, zhou2022covariance}. Many low-rank factor analysis techniques exist to improve the covariance matrix estimate. Here, we consider asset returns $\mathbf{x}$ as random variables that follow the model
\begin{equation} \label{Latent Model}
    \mathbf{x} = \mathbf{A} \mathbf{s} + \mathbf{e},
\end{equation}
where $\mathbf{x} \in \mathbb{R}^{n}$ is the observed data, $\mathbf{A} \in \mathbb{R}^{n \times r}$ is a factor loading matrix, $\mathbf{s} \in \mathbb{R}^{r}$ is the latent variable, and $\mathbf{e} \in \mathbb{R}^{n}$ is uncorrelated random noise, where $r \ll n$ \cite{stoica2023low}. Here, $\mathbf{s}$ represents macroeconomic factors like the growth rate of the GDP, inflation, unemployment, etc. We assume that $\mathbf{s}$ and $\mathbf{e}$ are uncorrelated and that data samples are independent and identically distributed. The covariance matrix is then $\mathbf{\Sigma} = \mathbb{E} [ \mathbf{x} \mathbf{x}^{\rm T} ]$. This gives
\begin{align}
    \mathbf{\Sigma} & = \mathbf{A} \mathbb{E} [ \mathbf{s} \mathbf{s}^{\rm T} ] \mathbf{A}^{\rm T} + \mathbb{E} [ \mathbf{e} \mathbf{e}^{\rm T} ] \\
     & = \mathbf{A} \mathbf{P} \mathbf{A}^{\rm T} + \mathbf{\Psi},
\end{align}
where $\mathbf{P} \equiv \mathbb{E} [ \mathbf{s} \mathbf{s}^{\rm T} ] \in \mathbb{R}^{r \times r}$ has ${\rm rank}(\mathbf{P}) \leq r$, ${\rm rank}(\mathbf{A}) \leq r$, and $\mathbf{\Psi}$ is a diagonal matrix containing the variance of noise on its diagonal
\begin{equation}
    \mathbf{\Psi} = \begin{bmatrix} \sigma_{1}^{2} & 0 & \cdots & 0 \\ 0 & \sigma_{2}^{2} & & 0 \\ \vdots & & \ddots & 0 \\ 0 & \cdots & 0 & \sigma_{n}^{2} \end{bmatrix}.
\end{equation}
Since ${\rm rank} ( \mathbf{AB} ) \leq \min ( {\rm rank} ( \mathbf{A} ), {\rm rank} ( \mathbf{B} ) )$, then ${\rm rank} ( \mathbf{A} \mathbf{P} \mathbf{A}^{\rm T} ) \leq r$. Therefore, we have decomposed the covariance matrix $\mathbf{\Sigma}$ into a positive semidefinite low-rank matrix plus a positive semidefinite diagonal matrix. Defining $\mathbf{M} \equiv \mathbf{A} \mathbf{P} \mathbf{A}^{\rm T}$, low-rank factor analysis concerns the estimation of $\mathbf{M}$ and $\mathbf{\Psi}$. To calculate $\mathbf{M}$ and $\mathbf{\Psi}$ we solve the minimization problem
\begin{equation} \label{Low-Rank Optimization}
\begin{aligned}
\min_{\mathbf{M}, \mathbf{\Psi}} \quad &  || \mathbf{S} - \mathbf{M} - \mathbf{\Psi} ||_{\rm F}^{2} \\
\textrm{s.t.} \quad & {\rm rank} (\mathbf{M}) \leq r, \\
& \mathbf{\Sigma} \succeq 0,
\end{aligned}
\end{equation}
where $|| \cdot ||_{\rm F}$ denotes the Frobenius norm \cite{manning2008introduction}. A common classical procedure for calculating matrix $\mathbf{M}$ in problem (\ref{Low-Rank Optimization}) using digital computers is given by the following steps:
\begin{enumerate}
    \item Construct the singular value decomposition (SVD) of $\mathbf{S}$. Since $\mathbf{S}$ is symmetric, its eigendecomposition is the same as the SVD, and we write $\mathbf{S} = \mathbf{U} \mathbf{\Lambda} \mathbf{U}^{\rm T}$, where $\mathbf{U}$ is the matrix of eigenvectors and $\mathbf{\Lambda}$ is the diagonal matrix of eigenvalues.
    \item Derive from $\mathbf{\Lambda}$ the matrix $\mathbf{\Lambda}_{r}$ formed by replacing with zeros the $n - r$ smallest eigenvalues on the diagonal of $\mathbf{\Lambda}$.
    \item Compute and output $\mathbf{M} = \mathbf{U} \mathbf{\Lambda}_{r} \mathbf{U}^{\rm T}$ as the rank-$r$ approximation to $\mathbf{S}$.
\end{enumerate}
Under the assumption $\mathbb{E} [ \mathbf{x} ] = \mathbf{0}$, the SVD method exactly replicates principal component analysis (PCA). The rank of $\mathbf{M}$ is at most $r$: this follows from the fact that $\mathbf{\Lambda}_r$ has at most $r$ non-zero values. Indeed, the Eckart-Young-Mirsky theorem proves that this procedure yields the matrix of rank less than or equal to $r$ with the lowest possible Frobenius error \cite{eckart1936approximation}. The diagonal matrix is estimated as $\mathbf{\Psi} = {\rm diag} (\mathbf{S} - \mathbf{M})$, where ${\rm diag} (\cdot)$ represents a diagonal matrix whose elements are $[ \mathbf{\Psi} ]_{ii} = [ \mathbf{S} - \mathbf{M} ]_{ii}$ and $[ \mathbf{\Psi} ]_{ij} = 0$ for $i \neq j$ \cite{bertsimas2017certifiably}. In addition, we constrain $[ \mathbf{\Psi} ]_{ii} \geq 0$, since the diagonal elements correspond to variances of the error variables. This guarantees that $\mathbf{\Sigma}$ is positive semidefinite. The eigendecomposition presented here becomes computationally expensive as the data size grows. Alternatively, autoencoders -- particularly when implemented using stochastic gradient descent -- can handle larger datasets and higher-dimensional data more efficiently than PCA \cite{kingma2013auto}. Additionally, when integrating dimensionality reduction as part of a larger neural network framework, an autoencoder can be easily embedded within the pipeline, whereas PCA would need to be applied as a separate pre-processing step \cite{lecun2015deep}.

\section{Linear Autoencoders}
\label{Linear Autoencoders}

\begin{figure}[t]
\centering
     \includegraphics[width=\columnwidth]{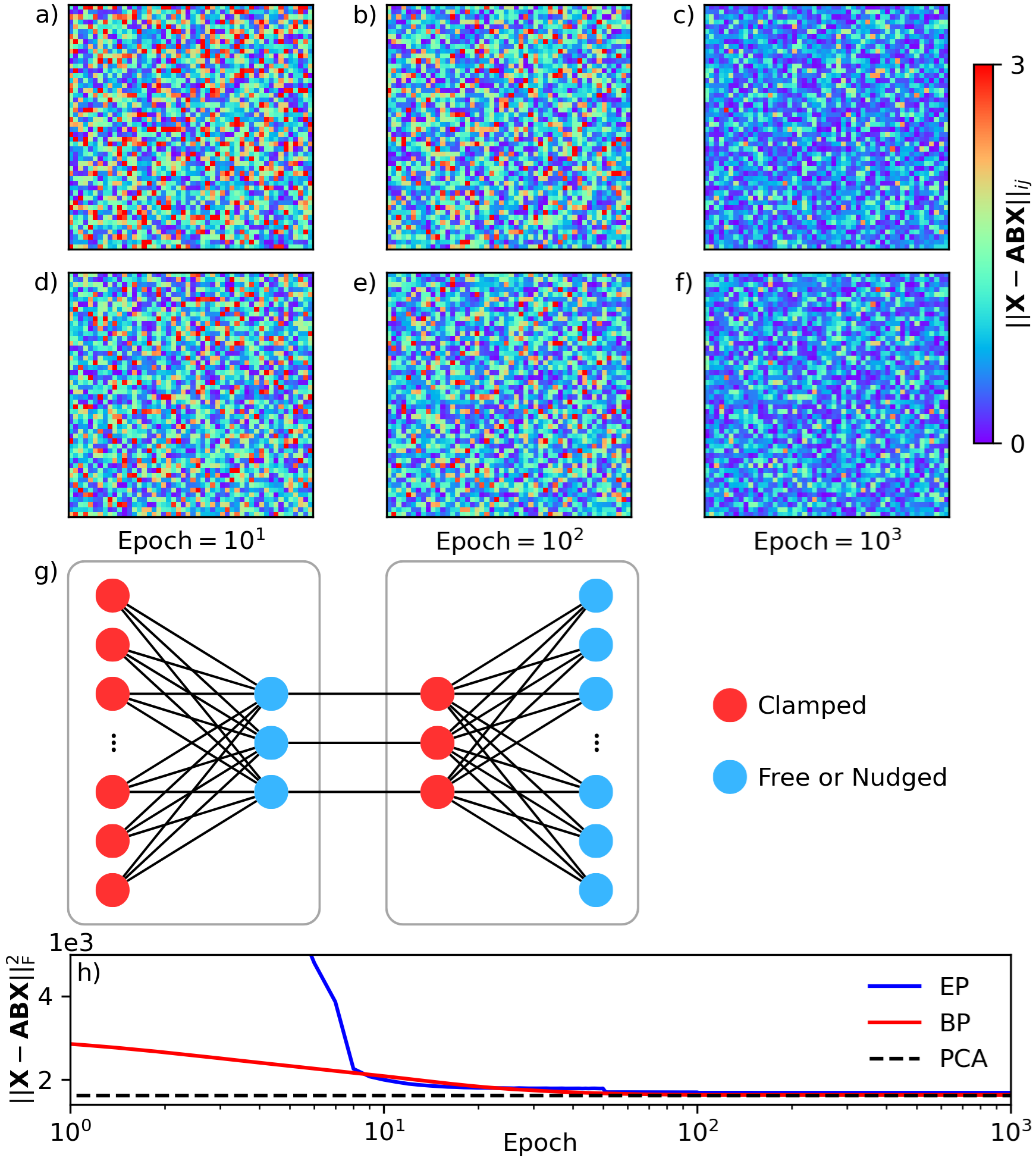}
     \caption{Training a linear autoencoder via (a)-(c) backpropagation (BP), and (d)-(f) with equilibrium propagation (EP). Input and output layers have size $50$, while the single hidden layer has size $5$. The networks are trained on $50$ vectors $\mathbf{x}_{1}, \ldots, \mathbf{x}_{50}$ of size $50$ whose elements are randomly sampled from the normal distribution $N (0, 1)$. (a)-(f) illustrate the element-wise absolute difference between the $50 \times 50$ matrix $\mathbf{X} = [\mathbf{x}_{1}, \ldots, \mathbf{x}_{50}]$ and its reconstructed output $\mathbf{ABX}$ at different epochs. (g) An illustrative example of the encoder/decoder Hopfield network structure trained with EP. (h) The overall network loss for BP and EP over epoch time. The black horizontal dashed line corresponds to the loss of the equivalent SVD/PCA method described in Section \ref{Low-Rank Approximation}.}
    \label{Linear Autoencoder Plot}
\end{figure}

A linear autoencoder is a classic neural network model for unsupervised learning that is trained to learn the identity function. The output and input layers  have the same number of nodes, while the middle layer has a fewer nodes. It aims to approximate the input through learning linear encodings and decodings between input and latent space. The encoder $\mathbf{B} \in \mathbb{R}^{r \times n}$ maps input $\mathbf{X} = [\mathbf{x}_{1}, \ldots, \mathbf{x}_{N}] \in \mathbb{R}^{n \times N}$ into a low-dimensional latent space $[\mathbf{s}_{1}, \ldots, \mathbf{s}_{N}]$, and the decoder $\mathbf{A} \in \mathbb{R}^{n \times r}$ maps $[\mathbf{s}_{1}, \ldots, \mathbf{s}_{N}]$ back to the original representation $\mathbf{X}$. We therefore recover the same model as in Eq.~(\ref{Latent Model}), and training the linear autoencoder becomes the minimization problem \cite{li2020rethink}
\begin{equation} \label{Autoencoder Norm}
\min_{\mathbf{A}, \mathbf{B}} \quad  || \mathbf{X} - \mathbf{A} \mathbf{B} \mathbf{X} ||_{\rm F}^{2}.
\end{equation}
We do not explicitly express the learnable biases in the network as these may be absorbed into the encoder $\mathbf{B}$ and decoder $\mathbf{A}$ by introducing an auxiliary row into $\mathbf{X}$ that is permanently clamped to values of $1$. We illustrate the training of a linear autoencoder in Fig.~(\ref{Linear Autoencoder Plot})(a)-(c) with the backpropagation method and compare it to the PCA in Fig.~(\ref{Linear Autoencoder Plot})(h). A linear autoencoder is related to the PCA. Indeed, under mild nondegeneracy conditions, any $\mathbf{A}$ at a local minimizer recovers the top rank-$r$ eigenspace of $\mathbf{X} \mathbf{X}^{\rm T}$ \cite{baldi1989neural}. However, unlike the actual PCA, the coordinates of the output of the middle layer in the network are correlated and are not sorted in descending order of variance \cite{plaut2018principal}. Autoencoder neural networks typically use backpropagation to train the weights. However, backpropagation is energy-intensive and not biologically plausible.

\section{Equilibrium Propagation}
\label{Equilibrium Propagation}

On dedicated analog hardware, equilibrium propagation is an energy-efficient alternative to backpropagation \cite{van2023training}. Therefore, in the supervised learning setting studied here, it may be used to train the weights of a linear autoencoder. Equilibrium propagation is an energy-based model because it relies on the concept of energy minimization to learn and make predictions. We consider the continuous Lyapunov function (\ref{HT_Energy}) with $p(t) = 1$ and $m_{i} = 0$ for all $i$, where here the symmetric coupling weights $J_{ij}$ are to be learned, and nonlinear activation function $g(x)$ need not be the same as in Section \ref{Continuous Hopfield Network}. Neurons $x_{i}$ are split in three sets: the input neurons, which are always clamped, the hidden neurons, and the output neurons. The discrepancy between the desired output $\mathbf{y}$ and the realized output $\hat{\mathbf{x}}$ is measured by the cost function
\begin{equation}
    C = \frac{1}{2} || \mathbf{y} - \hat{\mathbf{x}} ||_{2}^{2},
\end{equation}
which forms part of the total energy function $F = E + \beta C$. The clamping factor $\beta \geq 0$ is a  real-valued scalar that allows the output neurons to be weakly clamped \cite{scellier2017equilibrium}. The continuous-time dynamical system evolves according to the differential equation of motion
\begin{equation} \label{Equation of Motion}
    \frac{\md x_{i}}{\md t} = - \frac{\partial F}{\partial x_{i}} = - \frac{\partial E}{\partial x_{i}} - \beta \frac{\partial C}{\partial x_{i}},
\end{equation}
which is formed of two parts. The first is the internal force induced by the internal Hopfield energy, given by Eq.~(\ref{Hopfield}) for all $i$, and the second, the external force, is induced by the cost function $C$ as
\begin{equation} \label{Weakly Clamped Phase}
    - \beta \frac{\partial C}{\partial x_{i}} = \beta ( y_{i} - x_{i} ), \quad i \in \mathcal{Y},
\end{equation}
for nodes in the output layer $\mathcal{Y}$. Equilibrium propagation has two modes: the free phase and the weakly clamped phase. In the free phase $\beta = 0$ and only the inputs are clamped. The network then converges to a fixed point $\mathbf{x}^{*}$ and the output units are read out. In the weakly clamped phase $\beta > 0$, which induces an external force that acts on the output units as in Eq.~(\ref{Weakly Clamped Phase}). This force nudges the outputs from their fixed point values in the direction of the target values $y_{i}$. This perturbation propagates among the hidden neurons before a new fixed point $\mathbf{x}_{\beta}^{*}$ is found. Then, another weakly clamped phase is executed, this time with $\beta \to - \beta$ leading to the weakly clamped equilibrium $\mathbf{x}_{- \beta}^{*}$. It was shown that the weakly clamped phase implements the propagation of error derivatives with respect to the synaptic weights \cite{scellier2017equilibrium}. In the limit $\beta \to 0$, the update rule is
\begin{equation} \label{Update Rule}
    \Delta J_{ij} \propto \frac{1}{\beta} \left( \left. \frac{\partial F}{\partial J_{ij}} \right\rvert_{\mathbf{x}_{\beta}^{*}} - \left. \frac{\partial F}{\partial J_{ij}} \right\rvert_{\mathbf{x}_{- \beta}^{*}} \right),
\end{equation}
which is a second-order approximation to the standard backpropagation derivative \cite{laborieux2021scaling}. The process is iterated, at each step updating the weights $J_{ij}$ to minimize the loss function $C$. We choose activation function
\begin{equation} \label{Activation}
    g(x) = \begin{cases}
    x \quad & \text{if } | x | \leq c \\
    c \cdot \sgn (x) \quad & \text{otherwise},
    \end{cases}
\end{equation}
with constant $c$, so that under the condition that $| x_{i} | \leq c$ for all $i$, Eq.~(\ref{Hopfield}) is linear and can thus represent a linear autoencoder. In this case, the output $\hat{\mathbf{x}}$ of Eq.~(\ref{Hopfield}) is then the solution to the linear differential equation $\md \mathbf{x} / \md t = (\mathbf{J} - \mathbf{I}) \mathbf{x}$, and therefore
\begin{equation} \label{Steady State}
    \hat{\mathbf{x}} = \lim_{t \to \infty} \mathbf{x} (t) = \lim_{t \to \infty} \exp \{ (\mathbf{J} - \mathbf{I}) t \} \mathbf{x} (0).
\end{equation}
The constant $c$ in Eq.~(\ref{Activation}) is chosen to be large enough such that after training, all neurons obey $| x_{i} | \leq c$, and we can associate the Hopfield network as a linear autoencoder. To achieve a steady state in Eq.~(\ref{Steady State}), at least one eigenvalue of $\mathbf{J} - \mathbf{I}$ should be zero, with all others having a negative real part. 

\vspace{5pt}
\noindent \textbf{Proposition}. \textit{We state, with proof given in Ref.~\cite{baldi1989neural}, that for any fixed $n \times r$ matrix $\mathbf{A}$, Eq.~(\ref{Autoencoder Norm}) attains its minimum for $\mathbf{B} = (\mathbf{A}^{\rm T} \mathbf{A})^{-1} \mathbf{A}^{\rm T}$}.

\vspace{5pt}
\noindent \textbf{Lemma}. \textit{The $n \times n$ matrix $\mathbf{J} - \mathbf{I}$, where $\mathbf{J} = \mathbf{AB}$, has at least one zero eigenvalue, with all others having negative real part}.

\vspace{5pt}
\noindent \textit{Proof}. $\mathbf{J} = \mathbf{AB} = \mathbf{A} (\mathbf{A}^{\rm T} \mathbf{A})^{-1} \mathbf{A}^{\rm T}$, and therefore
\begin{align}
    \mathbf{J}^{2} & = \mathbf{A} (\mathbf{A}^{\rm T} \mathbf{A})^{-1} \mathbf{A}^{\rm T} \mathbf{A} (\mathbf{A}^{\rm T} \mathbf{A})^{-1} \mathbf{A}^{\rm T} \\
     & = \mathbf{A} (\mathbf{A}^{\rm T} \mathbf{A})^{-1} \mathbf{A}^{\rm T},
\end{align}
which shows that $\mathbf{J}$ is idempotent, that is $\mathbf{J}^{2} = \mathbf{J}$. It follows that $\mathbf{J}$ is a projection operator on the column space $C (\mathbf{J})$ along its null space $N (\mathbf{J})$. The $n$ eigenvalues $\lambda_{i}$ of $\mathbf{J}$ are either $0$ or $1$: $\lambda_{i} \mathbf{x}_{i} = \mathbf{J} \mathbf{x}_{i} = \mathbf{J}^{2} \mathbf{x}_{i} = \lambda_{i} \mathbf{J} \mathbf{x}_{i} = \lambda_{i}^{2} \mathbf{x}_{i}$, which implies $\lambda_{i} \in \{ 0, 1 \}$. By construction, $\mathbf{J}$ has rank at most $r$, and therefore there are at least $n - r$ zero eigenvalues. It follows that there are between $1$ and $r$ nonzero eigenvalues of $\mathbf{J}$, which must have value $\lambda_{i} = 1$. Since $\mathbf{J} - \mathbf{I}$ has eigenvalues $\mu_{i} = \lambda_{i} - 1$, then $\mu_i \in \{ -1, 0 \}$. Therefore, $\mathbf{J} - \mathbf{I}$ has between $1$ and $r$ zero eigenvalues, with all others being equal to $-1$.
\vspace{5pt}

The Lemma guarantees that should equilibrium propagation learn the weights that minimize Eq.~(\ref{Autoencoder Norm}), the corresponding Hopfield network will converge to a steady state. Yet, during training, this will, in general, not be the case, and positive eigenvalues of $\mathbf{J} - \mathbf{I}$ will produce exponential growth in Eq.~(\ref{Steady State}). However, Eq.~(\ref{Steady State}) only holds in the linear regime of the activation function (\ref{Activation}). Exponential growth is prevented by the symmetric clipping incorporated into the nonlinear activation function $g(x)$ for neurons with $| x_{i} | > c$.

In the linear regime, the overall network dynamics is represented by the square matrix $\lim_{t \to \infty} \exp \{ (\mathbf{J} - \mathbf{I}) t \}$, which for linear autoencoders we seek to decompose into its non-square constituent parts: encoder $\mathbf{B}$ and decoder $\mathbf{A}$. We achieve this by treating the encoder and decoder as separate Hopfield networks, as shown in Fig.~(\ref{Linear Autoencoder Plot})(g), each with their own energy function. The encoder settles into an equilibrium representing the latent vector $\mathbf{s}$ without taking into account the decoder. $\mathbf{s}$ is then used as a fixed input to the decoder which then settles into its own equilibrium. The decoder then undergoes the weakly clamped phases, and its weights are updated according to Eq.~(\ref{Update Rule}). The encoder weights also need to be optimized to lower the reconstruction loss at the decoder output, which is achieved by setting
\begin{equation} \label{Encoder Update}
    \frac{\partial C}{\partial x_{i}} = \lim_{\beta \to 0} \frac{1}{2 \beta} \left( \left. \frac{\partial F}{\partial x_{i}} \right\rvert_{\mathbf{x}_{\beta}^{\rm (dec)}} - \left. \frac{\partial F}{\partial x_{i}} \right\rvert_{\mathbf{x}_{- \beta}^{\rm (dec)}} \right), \quad i \in \mathcal{Y},
\end{equation}
in Eq.~(\ref{Equation of Motion}), where $\mathbf{x}_{\beta}^{\rm (dec)}$ is the weakly clamped decoder equilibrium state, and Eq.~(\ref{Encoder Update}) only pertains to neurons in the encoder output layer $\mathcal{Y}$. Equation (\ref{Encoder Update}) follows from the fact that it can be shown that equilibrium propagation also allows for finding the gradient of the loss with respect to the input \cite{scellier2021deep}. We note that $\mathbf{J}$, which contains the couplings of the continuous Hopfield network, is now a $(n + r) \times (n + r)$ matrix on account of the number of nodes in the encoder and decoder networks. Nonetheless, the factor loading matrix $\mathbf{A}$ can be recovered as the $n \times r$ block corresponding to the nodes of the decoder output layer. The equilibrium propagation training procedure is illustrated in Fig.~(\ref{Linear Autoencoder Plot})(d)-(f) and compared to backpropagation and the PCA in Fig.~(\ref{Linear Autoencoder Plot})(h).

\section{Results}
\label{Results}

\begin{figure*}[t]
\centering
     \includegraphics[width=2\columnwidth]{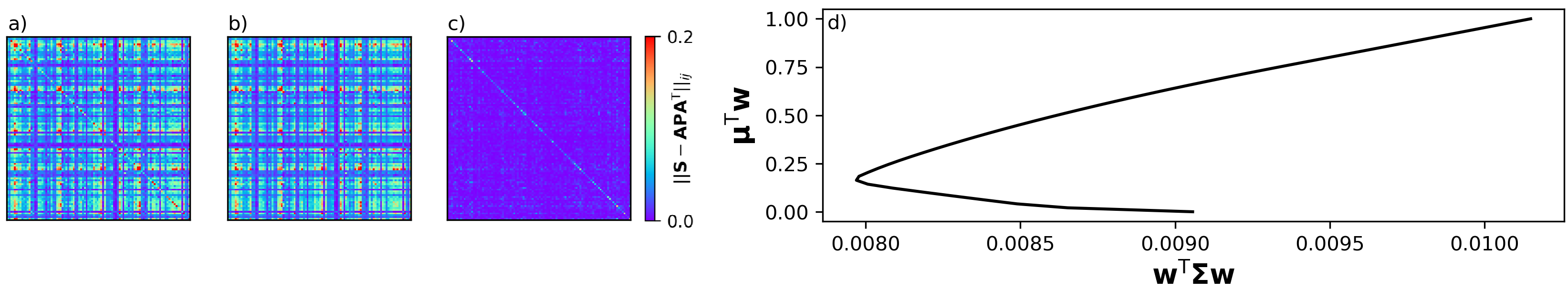}
     \caption{(a) The sample covariance matrix $\mathbf{S}$ for $n = 100$ financial stocks selected from the S\&P 500 index. $\mathbf{S}$ is calculated from Eq.~(\ref{Sample Covariance}), with $N = 50$ time series samples. (b) The $r = 10$ low-rank approximation $\mathbf{APA}^{\rm T}$ of the covariance matrix, as calculated by training a continuous Hopfield network via equilibrium propagation. (c) The element-wise absolute difference between the sample covariance matrix and its low-rank approximation. (d) The hyperbola in variance-return space for possible portfolios. Each point along the hyperbola is calculated by solving (\ref{Portfolio Optimization}) for a specific return value $R$ using Eq.~(\ref{Hopfield}).}
    \label{Results Plot}
\end{figure*}

We present our results in an order consistent with the portfolio optimization process a company or institution may take. We start with raw data samples and construct a low-rank covariance matrix approximating the true covariance matrix. This is a common procedure because the number of observations is often smaller than the number of assets, leading to significant noise that can distort the underlying relationships between the assets. Next, we calculate the efficient frontier, which produces optimal portfolios for each level of desired expected return $R$.

We begin by collecting real data samples $\mathbf{x}_{i} \in \mathbb{R}^{n}$ for $i = 1, 2, \ldots, N$ from stock returns of a selection of $n = 100$ stocks in the S\&P 500 index. We restrict ourselves to only $N = 50$ observations such that the sample covariance matrix has a tendency to contain significant noise. Two continuous Hopfield networks, structured as the encoder and decoder parts of a linear autoencoder, are trained using equilibrium propagation. The latent variables $[ \mathbf{s}_{1}, \ldots, \mathbf{s}_{N}]$ are calculated as the subset $\mathcal{Y}$ of steady-state solutions of the encoder network, while the factor loading matrix $\mathbf{A}$ is the $n \times r$ block of the decoder matrix representation $\lim_{t \to \infty} \exp \{ ( \mathbf{J} - \mathbf{I} ) t \}$ corresponding to its output layer $\mathcal{Y}$. In practice, we cannot take the limit to infinity, and instead, we use a suitably large value of $t$ such that $\exp \{ ( \mathbf{J} - \mathbf{I} ) t \}$ changes minimally from $t$ to $t + 1$. The low-rank approximation to the covariance matrix is calculated as $\mathbf{A} \mathbf{P} \mathbf{A}^{\rm T}$, where $\mathbf{P} = \mathbb{E} [ \mathbf{s} \mathbf{s}^{\rm T} ]$. We depict the full-rank sample covariance matrix and the equilibrium propagation-based low-rank approximation in Figs.~(\ref{Results Plot})(a) and (b) respectively. Figure (\ref{Results Plot})(c) then illustrates the element-wise absolute difference between these two covariance matrices. The low-rank approximation is plugged into (\ref{Portfolio Optimization}) and solved for the portfolio weights $\mathbf{w}$ using the continuous Hopfield network of Eq.~(\ref{Hopfield}). We minimize the portfolio variance subject to the constraint $\boldsymbol{\mu}^{\rm T} \mathbf{w} = R$ for incremental values of $R$. In Fig.~(\ref{Results Plot})(d), we plot the corresponding variances and returns for range $R = [0, 1]$. The efficient frontier is identified, and an optimal portfolio can be selected based on risk appetite. In particular, the minimum variance and maximum Sharpe ratio portfolios can be established.

Analog Hopfield networks can be implemented as electronic circuits \cite{cai2020power} and photonic neural networks \cite{tait2017neuromorphic}. Photonic systems operate on picosecond to femtosecond timescales as high bandwidth signals flow through a single optical waveguide. Consequently, such implementations can have dense connectivity while maintaining fast convergence times. However, physical analog platforms are subject to noise sensitivity, thermal effects, and non-idealities in circuit components which can degrade performance. In addition, real-world portfolio optimization problems often involve complex constraints such as transaction costs, market liquidity, regulatory requirements, cardinality constraints, and tax considerations. While some of these can be readily incorporated into the objective function (\ref{Equivalent}), for example, an $\ell^{1}$-norm can enforce sparsity to satisfy a cardinality constraint, others take more complex forms. For instance, a hybrid approach that combines analog Hopfield networks with digital computing could be explored. 

\section{Conclusions}

This paper introduces a fully analog pipeline for portfolio optimization problems. Starting with raw data samples, the proposed pipeline leverages the energy-efficient analog operation of continuous Hopfield networks to calculate optimal portfolio weights. The analog pipeline distinguishes itself from traditional digital methods by its speed and scalability, with applications in time-sensitive domains such as high-frequency trading. At the heart of the pipeline are continuous Hopfield networks, used in two separate applications: autoencoder neural networks and minimum variance portfolios.

The autoencoder network yields the latent variables and factor loading matrix, which are then used to calculate a low-rank approximation of the covariance matrix. After that,  a Hopfield network applies the quadratic mean-variance model to determine optimal portfolio weights.
By shifting to analog architectures, we reduce the reliance on binary logic operations typical of digital systems, paving the way for a more energy-efficient approach to computation. This efficiency can reduce power consumption in data centers and other computing environments, addressing the growing energy demands of digital computing. Specifically, companies can reduce their energy consumption while optimizing large portfolios as part of their risk management processes.

\section*{Acknowledgments}

J.S.C.\ acknowledges the PhD support from the EPSRC. N.G.B.\ acknowledges support from HORIZON EIC-2022-PATHFINDERCHALLENGES-01 HEISINGBERG Project 101114978 and Weizmann-UK Make Connection Grant 142568.

\bibliography{references}

\begin{thebibliography}{35}%
\makeatletter
\providecommand \@ifxundefined [1]{%
 \@ifx{#1\undefined}
}%
\providecommand \@ifnum [1]{%
 \ifnum #1\expandafter \@firstoftwo
 \else \expandafter \@secondoftwo
 \fi
}%
\providecommand \@ifx [1]{%
 \ifx #1\expandafter \@firstoftwo
 \else \expandafter \@secondoftwo
 \fi
}%
\providecommand \natexlab [1]{#1}%
\providecommand \enquote  [1]{``#1''}%
\providecommand \bibnamefont  [1]{#1}%
\providecommand \bibfnamefont [1]{#1}%
\providecommand \citenamefont [1]{#1}%
\providecommand \href@noop [0]{\@secondoftwo}%
\providecommand \href [0]{\begingroup \@sanitize@url \@href}%
\providecommand \@href[1]{\@@startlink{#1}\@@href}%
\providecommand \@@href[1]{\endgroup#1\@@endlink}%
\providecommand \@sanitize@url [0]{\catcode `\\12\catcode `\$12\catcode
  `\&12\catcode `\#12\catcode `\^12\catcode `\_12\catcode `\%12\relax}%
\providecommand \@@startlink[1]{}%
\providecommand \@@endlink[0]{}%
\providecommand \url  [0]{\begingroup\@sanitize@url \@url }%
\providecommand \@url [1]{\endgroup\@href {#1}{\urlprefix }}%
\providecommand \urlprefix  [0]{URL }%
\providecommand \Eprint [0]{\href }%
\providecommand \doibase [0]{http://dx.doi.org/}%
\providecommand \selectlanguage [0]{\@gobble}%
\providecommand \bibinfo  [0]{\@secondoftwo}%
\providecommand \bibfield  [0]{\@secondoftwo}%
\providecommand \translation [1]{[#1]}%
\providecommand \BibitemOpen [0]{}%
\providecommand \bibitemStop [0]{}%
\providecommand \bibitemNoStop [0]{.\EOS\space}%
\providecommand \EOS [0]{\spacefactor3000\relax}%
\providecommand \BibitemShut  [1]{\csname bibitem#1\endcsname}%
\let\auto@bib@innerbib\@empty
\bibitem [{\citenamefont {Markowitz}(1952)}]{markowitz1952portfolio}%
  \BibitemOpen
  \bibfield  {author} {\bibinfo {author} {\bibfnamefont {H.}~\bibnamefont
  {Markowitz}},\ }\href@noop {} {\bibfield  {journal} {\bibinfo  {journal} {The
  Journal of Finance}\ }\textbf {\bibinfo {volume} {7}},\ \bibinfo {pages} {77}
  (\bibinfo {year} {1952})}\BibitemShut {NoStop}%
\bibitem [{\citenamefont {Ledoit}\ and\ \citenamefont
  {Wolf}(2022)}]{ledoit2022power}%
  \BibitemOpen
  \bibfield  {author} {\bibinfo {author} {\bibfnamefont {O.}~\bibnamefont
  {Ledoit}}\ and\ \bibinfo {author} {\bibfnamefont {M.}~\bibnamefont {Wolf}},\
  }\href@noop {} {\bibfield  {journal} {\bibinfo  {journal} {Journal of
  Financial Econometrics}\ }\textbf {\bibinfo {volume} {20}},\ \bibinfo {pages}
  {187} (\bibinfo {year} {2022})}\BibitemShut {NoStop}%
\bibitem [{\citenamefont {Tabeart}\ \emph {et~al.}(2020)\citenamefont
  {Tabeart}, \citenamefont {Dance}, \citenamefont {Lawless}, \citenamefont
  {Nichols},\ and\ \citenamefont {Waller}}]{tabeart2020improving}%
  \BibitemOpen
  \bibfield  {author} {\bibinfo {author} {\bibfnamefont {J.~M.}\ \bibnamefont
  {Tabeart}}, \bibinfo {author} {\bibfnamefont {S.~L.}\ \bibnamefont {Dance}},
  \bibinfo {author} {\bibfnamefont {A.~S.}\ \bibnamefont {Lawless}}, \bibinfo
  {author} {\bibfnamefont {N.~K.}\ \bibnamefont {Nichols}}, \ and\ \bibinfo
  {author} {\bibfnamefont {J.~A.}\ \bibnamefont {Waller}},\ }\href@noop {}
  {\bibfield  {journal} {\bibinfo  {journal} {Tellus A: Dynamic Meteorology and
  Oceanography}\ }\textbf {\bibinfo {volume} {72}},\ \bibinfo {pages} {1}
  (\bibinfo {year} {2020})}\BibitemShut {NoStop}%
\bibitem [{\citenamefont {Bai}\ and\ \citenamefont
  {Ng}(2002)}]{bai2002determining}%
  \BibitemOpen
  \bibfield  {author} {\bibinfo {author} {\bibfnamefont {J.}~\bibnamefont
  {Bai}}\ and\ \bibinfo {author} {\bibfnamefont {S.}~\bibnamefont {Ng}},\
  }\href@noop {} {\bibfield  {journal} {\bibinfo  {journal} {Econometrica}\
  }\textbf {\bibinfo {volume} {70}},\ \bibinfo {pages} {191} (\bibinfo {year}
  {2002})}\BibitemShut {NoStop}%
\bibitem [{\citenamefont {Liu}(2009)}]{liu2009portfolio}%
  \BibitemOpen
  \bibfield  {author} {\bibinfo {author} {\bibfnamefont {Q.}~\bibnamefont
  {Liu}},\ }\href@noop {} {\bibfield  {journal} {\bibinfo  {journal} {Journal
  of Applied Econometrics}\ }\textbf {\bibinfo {volume} {24}},\ \bibinfo
  {pages} {560} (\bibinfo {year} {2009})}\BibitemShut {NoStop}%
\bibitem [{\citenamefont {Goumatianos}\ \emph {et~al.}(2013)\citenamefont
  {Goumatianos}, \citenamefont {Christou},\ and\ \citenamefont
  {Lindgren}}]{goumatianos2013stock}%
  \BibitemOpen
  \bibfield  {author} {\bibinfo {author} {\bibfnamefont {N.}~\bibnamefont
  {Goumatianos}}, \bibinfo {author} {\bibfnamefont {I.}~\bibnamefont
  {Christou}}, \ and\ \bibinfo {author} {\bibfnamefont {P.}~\bibnamefont
  {Lindgren}},\ }\href@noop {} {\bibfield  {journal} {\bibinfo  {journal}
  {Procedia Economics and Finance}\ }\textbf {\bibinfo {volume} {5}},\ \bibinfo
  {pages} {298} (\bibinfo {year} {2013})}\BibitemShut {NoStop}%
\bibitem [{\citenamefont {Ziegelmann}\ \emph {et~al.}(2015)\citenamefont
  {Ziegelmann}, \citenamefont {Borges},\ and\ \citenamefont
  {Caldeira}}]{ziegelmann2015selection}%
  \BibitemOpen
  \bibfield  {author} {\bibinfo {author} {\bibfnamefont {F.~A.}\ \bibnamefont
  {Ziegelmann}}, \bibinfo {author} {\bibfnamefont {B.}~\bibnamefont {Borges}},
  \ and\ \bibinfo {author} {\bibfnamefont {J.~F.}\ \bibnamefont {Caldeira}},\
  }\href@noop {} {\bibfield  {journal} {\bibinfo  {journal} {Brazilian Review
  of Econometrics}\ }\textbf {\bibinfo {volume} {35}},\ \bibinfo {pages} {23}
  (\bibinfo {year} {2015})}\BibitemShut {NoStop}%
\bibitem [{\citenamefont {Filipiak}\ and\ \citenamefont
  {Lipinski}(2017)}]{filipiak2017dynamic}%
  \BibitemOpen
  \bibfield  {author} {\bibinfo {author} {\bibfnamefont {P.}~\bibnamefont
  {Filipiak}}\ and\ \bibinfo {author} {\bibfnamefont {P.}~\bibnamefont
  {Lipinski}},\ }in\ \href@noop {} {\emph {\bibinfo {booktitle} {Applications
  of Evolutionary Computation: 20th European Conference, Proceedings, Part I
  20}}}\ (\bibinfo {organization} {Springer},\ \bibinfo {year} {2017})\ pp.\
  \bibinfo {pages} {34--50}\BibitemShut {NoStop}%
\bibitem [{\citenamefont {Vadlamani}\ \emph {et~al.}(2020)\citenamefont
  {Vadlamani}, \citenamefont {Xiao},\ and\ \citenamefont
  {Yablonovitch}}]{vadlamani2020physics}%
  \BibitemOpen
  \bibfield  {author} {\bibinfo {author} {\bibfnamefont {S.~K.}\ \bibnamefont
  {Vadlamani}}, \bibinfo {author} {\bibfnamefont {T.~P.}\ \bibnamefont {Xiao}},
  \ and\ \bibinfo {author} {\bibfnamefont {E.}~\bibnamefont {Yablonovitch}},\
  }\href@noop {} {\bibfield  {journal} {\bibinfo  {journal} {Proceedings of the
  National Academy of Sciences}\ }\textbf {\bibinfo {volume} {117}},\ \bibinfo
  {pages} {26639} (\bibinfo {year} {2020})}\BibitemShut {NoStop}%
\bibitem [{\citenamefont {Farhi}\ \emph {et~al.}(2000)\citenamefont {Farhi},
  \citenamefont {Goldstone}, \citenamefont {Gutmann},\ and\ \citenamefont
  {Sipser}}]{farhi}%
  \BibitemOpen
  \bibfield  {author} {\bibinfo {author} {\bibfnamefont {E.}~\bibnamefont
  {Farhi}}, \bibinfo {author} {\bibfnamefont {J.}~\bibnamefont {Goldstone}},
  \bibinfo {author} {\bibfnamefont {S.}~\bibnamefont {Gutmann}}, \ and\
  \bibinfo {author} {\bibfnamefont {M.}~\bibnamefont {Sipser}},\ }\href@noop {}
  {\bibfield  {journal} {\bibinfo  {journal} {arXiv preprint quant-ph/0001106}\
  } (\bibinfo {year} {2000})}\BibitemShut {NoStop}%
\bibitem [{\citenamefont {Lucas}(2014)}]{lucas2014ising}%
  \BibitemOpen
  \bibfield  {author} {\bibinfo {author} {\bibfnamefont {A.}~\bibnamefont
  {Lucas}},\ }\href@noop {} {\bibfield  {journal} {\bibinfo  {journal}
  {Frontiers in Physics}\ }\textbf {\bibinfo {volume} {2}},\ \bibinfo {pages}
  {5} (\bibinfo {year} {2014})}\BibitemShut {NoStop}%
\bibitem [{\citenamefont {Berloff}\ \emph {et~al.}(2017)\citenamefont
  {Berloff}, \citenamefont {Silva}, \citenamefont {Kalinin}, \citenamefont
  {Askitopoulos}, \citenamefont {T{\"o}pfer}, \citenamefont {Cilibrizzi},
  \citenamefont {Langbein},\ and\ \citenamefont
  {Lagoudakis}}]{berloff2017realizing}%
  \BibitemOpen
  \bibfield  {author} {\bibinfo {author} {\bibfnamefont {N.~G.}\ \bibnamefont
  {Berloff}}, \bibinfo {author} {\bibfnamefont {M.}~\bibnamefont {Silva}},
  \bibinfo {author} {\bibfnamefont {K.}~\bibnamefont {Kalinin}}, \bibinfo
  {author} {\bibfnamefont {A.}~\bibnamefont {Askitopoulos}}, \bibinfo {author}
  {\bibfnamefont {J.~D.}\ \bibnamefont {T{\"o}pfer}}, \bibinfo {author}
  {\bibfnamefont {P.}~\bibnamefont {Cilibrizzi}}, \bibinfo {author}
  {\bibfnamefont {W.}~\bibnamefont {Langbein}}, \ and\ \bibinfo {author}
  {\bibfnamefont {P.~G.}\ \bibnamefont {Lagoudakis}},\ }\href@noop {}
  {\bibfield  {journal} {\bibinfo  {journal} {Nature Materials}\ } (\bibinfo
  {year} {2017})}\BibitemShut {NoStop}%
\bibitem [{\citenamefont {Kalinin}\ \emph {et~al.}(2020)\citenamefont
  {Kalinin}, \citenamefont {Amo}, \citenamefont {Bloch},\ and\ \citenamefont
  {Berloff}}]{kalinin2020polaritonic}%
  \BibitemOpen
  \bibfield  {author} {\bibinfo {author} {\bibfnamefont {K.~P.}\ \bibnamefont
  {Kalinin}}, \bibinfo {author} {\bibfnamefont {A.}~\bibnamefont {Amo}},
  \bibinfo {author} {\bibfnamefont {J.}~\bibnamefont {Bloch}}, \ and\ \bibinfo
  {author} {\bibfnamefont {N.~G.}\ \bibnamefont {Berloff}},\ }\href@noop {}
  {\bibfield  {journal} {\bibinfo  {journal} {Nanophotonics}\ }\textbf
  {\bibinfo {volume} {9}},\ \bibinfo {pages} {4127} (\bibinfo {year}
  {2020})}\BibitemShut {NoStop}%
\bibitem [{\citenamefont {Beasley}(2013)}]{beasley2013portfolio}%
  \BibitemOpen
  \bibfield  {author} {\bibinfo {author} {\bibfnamefont {J.~E.}\ \bibnamefont
  {Beasley}},\ }in\ \href@noop {} {\emph {\bibinfo {booktitle} {Theory Driven
  by Influential Applications}}}\ (\bibinfo  {publisher} {Informs},\ \bibinfo
  {year} {2013})\ pp.\ \bibinfo {pages} {201--221}\BibitemShut {NoStop}%
\bibitem [{\citenamefont {Wang}\ \emph {et~al.}(2024)\citenamefont {Wang},
  \citenamefont {Cummins}, \citenamefont {Syed}, \citenamefont {Stroev},
  \citenamefont {Pastras}, \citenamefont {Sakellariou}, \citenamefont
  {Tsintzos}, \citenamefont {Askitopoulos}, \citenamefont {Veraldi},
  \citenamefont {Strinati} \emph {et~al.}}]{wang2024efficient}%
  \BibitemOpen
  \bibfield  {author} {\bibinfo {author} {\bibfnamefont {R.~Z.}\ \bibnamefont
  {Wang}}, \bibinfo {author} {\bibfnamefont {J.~S.}\ \bibnamefont {Cummins}},
  \bibinfo {author} {\bibfnamefont {M.}~\bibnamefont {Syed}}, \bibinfo {author}
  {\bibfnamefont {N.}~\bibnamefont {Stroev}}, \bibinfo {author} {\bibfnamefont
  {G.}~\bibnamefont {Pastras}}, \bibinfo {author} {\bibfnamefont
  {J.}~\bibnamefont {Sakellariou}}, \bibinfo {author} {\bibfnamefont
  {S.}~\bibnamefont {Tsintzos}}, \bibinfo {author} {\bibfnamefont
  {A.}~\bibnamefont {Askitopoulos}}, \bibinfo {author} {\bibfnamefont
  {D.}~\bibnamefont {Veraldi}}, \bibinfo {author} {\bibfnamefont {M.~C.}\
  \bibnamefont {Strinati}},  \emph {et~al.},\ }\href@noop {} {\bibfield
  {journal} {\bibinfo  {journal} {arXiv preprint arXiv:2406.01400}\ } (\bibinfo
  {year} {2024})}\BibitemShut {NoStop}%
\bibitem [{\citenamefont {De~las Cuevas}\ and\ \citenamefont
  {Cubitt}(2016)}]{Cubitt_universality}%
  \BibitemOpen
  \bibfield  {author} {\bibinfo {author} {\bibfnamefont {G.}~\bibnamefont
  {De~las Cuevas}}\ and\ \bibinfo {author} {\bibfnamefont {T.~S.}\ \bibnamefont
  {Cubitt}},\ }\href@noop {} {\bibfield  {journal} {\bibinfo  {journal}
  {Science}\ }\textbf {\bibinfo {volume} {351}},\ \bibinfo {pages} {1180}
  (\bibinfo {year} {2016})}\BibitemShut {NoStop}%
\bibitem [{\citenamefont {Hopfield}(1984)}]{hopfield1984neurons}%
  \BibitemOpen
  \bibfield  {author} {\bibinfo {author} {\bibfnamefont {J.~J.}\ \bibnamefont
  {Hopfield}},\ }\href@noop {} {\bibfield  {journal} {\bibinfo  {journal}
  {Proceedings of the National Academy of Sciences}\ }\textbf {\bibinfo
  {volume} {81}},\ \bibinfo {pages} {3088} (\bibinfo {year}
  {1984})}\BibitemShut {NoStop}%
\bibitem [{\citenamefont {Kalinin}\ \emph {et~al.}(2023)\citenamefont
  {Kalinin}, \citenamefont {Mourgias-Alexandris}, \citenamefont {Ballani},
  \citenamefont {Berloff}, \citenamefont {Clegg}, \citenamefont {Cletheroe},
  \citenamefont {Gkantsidis}, \citenamefont {Haller}, \citenamefont
  {Lyutsarev}, \citenamefont {Parmigiani}, \citenamefont {Pickup} \emph
  {et~al.}}]{kalinin2023analog}%
  \BibitemOpen
  \bibfield  {author} {\bibinfo {author} {\bibfnamefont {K.}~\bibnamefont
  {Kalinin}}, \bibinfo {author} {\bibfnamefont {G.}~\bibnamefont
  {Mourgias-Alexandris}}, \bibinfo {author} {\bibfnamefont {H.}~\bibnamefont
  {Ballani}}, \bibinfo {author} {\bibfnamefont {N.~G.}\ \bibnamefont
  {Berloff}}, \bibinfo {author} {\bibfnamefont {J.~H.}\ \bibnamefont {Clegg}},
  \bibinfo {author} {\bibfnamefont {D.}~\bibnamefont {Cletheroe}}, \bibinfo
  {author} {\bibfnamefont {C.}~\bibnamefont {Gkantsidis}}, \bibinfo {author}
  {\bibfnamefont {I.}~\bibnamefont {Haller}}, \bibinfo {author} {\bibfnamefont
  {V.}~\bibnamefont {Lyutsarev}}, \bibinfo {author} {\bibfnamefont
  {F.}~\bibnamefont {Parmigiani}}, \bibinfo {author} {\bibfnamefont
  {L.}~\bibnamefont {Pickup}},  \emph {et~al.},\ }\href@noop {} {\bibfield
  {journal} {\bibinfo  {journal} {arXiv preprint arXiv:2304.12594}\ } (\bibinfo
  {year} {2023})}\BibitemShut {NoStop}%
\bibitem [{\citenamefont {Goto}(2016)}]{goto2016bifurcation}%
  \BibitemOpen
  \bibfield  {author} {\bibinfo {author} {\bibfnamefont {H.}~\bibnamefont
  {Goto}},\ }\href@noop {} {\bibfield  {journal} {\bibinfo  {journal}
  {Scientific Reports}\ }\textbf {\bibinfo {volume} {6}},\ \bibinfo {pages} {1}
  (\bibinfo {year} {2016})}\BibitemShut {NoStop}%
\bibitem [{\citenamefont {Zhou}\ \emph {et~al.}(2022)\citenamefont {Zhou},
  \citenamefont {Ying},\ and\ \citenamefont {Palomar}}]{zhou2022covariance}%
  \BibitemOpen
  \bibfield  {author} {\bibinfo {author} {\bibfnamefont {R.}~\bibnamefont
  {Zhou}}, \bibinfo {author} {\bibfnamefont {J.}~\bibnamefont {Ying}}, \ and\
  \bibinfo {author} {\bibfnamefont {D.~P.}\ \bibnamefont {Palomar}},\
  }\href@noop {} {\bibfield  {journal} {\bibinfo  {journal} {IEEE Transactions
  on Signal Processing}\ }\textbf {\bibinfo {volume} {70}},\ \bibinfo {pages}
  {4020} (\bibinfo {year} {2022})}\BibitemShut {NoStop}%
\bibitem [{\citenamefont {Stoica}\ and\ \citenamefont
  {Babu}(2023)}]{stoica2023low}%
  \BibitemOpen
  \bibfield  {author} {\bibinfo {author} {\bibfnamefont {P.}~\bibnamefont
  {Stoica}}\ and\ \bibinfo {author} {\bibfnamefont {P.}~\bibnamefont {Babu}},\
  }\href@noop {} {\bibfield  {journal} {\bibinfo  {journal} {IEEE Transactions
  on Signal Processing}\ }\textbf {\bibinfo {volume} {71}},\ \bibinfo {pages}
  {1699} (\bibinfo {year} {2023})}\BibitemShut {NoStop}%
\bibitem [{\citenamefont {Manning}(2008)}]{manning2008introduction}%
  \BibitemOpen
  \bibfield  {author} {\bibinfo {author} {\bibfnamefont {C.~D.}\ \bibnamefont
  {Manning}},\ }\href@noop {} {\emph {\bibinfo {title} {Introduction to
  Information Retrieval}}}\ (\bibinfo  {publisher} {Syngress Publishing},\
  \bibinfo {year} {2008})\BibitemShut {NoStop}%
\bibitem [{\citenamefont {Eckart}\ and\ \citenamefont
  {Young}(1936)}]{eckart1936approximation}%
  \BibitemOpen
  \bibfield  {author} {\bibinfo {author} {\bibfnamefont {C.}~\bibnamefont
  {Eckart}}\ and\ \bibinfo {author} {\bibfnamefont {G.}~\bibnamefont {Young}},\
  }\href@noop {} {\bibfield  {journal} {\bibinfo  {journal} {Psychometrika}\
  }\textbf {\bibinfo {volume} {1}},\ \bibinfo {pages} {211} (\bibinfo {year}
  {1936})}\BibitemShut {NoStop}%
\bibitem [{\citenamefont {Bertsimas}\ \emph {et~al.}(2017)\citenamefont
  {Bertsimas}, \citenamefont {Copenhaver},\ and\ \citenamefont
  {Mazumder}}]{bertsimas2017certifiably}%
  \BibitemOpen
  \bibfield  {author} {\bibinfo {author} {\bibfnamefont {D.}~\bibnamefont
  {Bertsimas}}, \bibinfo {author} {\bibfnamefont {M.~S.}\ \bibnamefont
  {Copenhaver}}, \ and\ \bibinfo {author} {\bibfnamefont {R.}~\bibnamefont
  {Mazumder}},\ }\href@noop {} {\bibfield  {journal} {\bibinfo  {journal}
  {Journal of Machine Learning Research}\ }\textbf {\bibinfo {volume} {18}},\
  \bibinfo {pages} {1} (\bibinfo {year} {2017})}\BibitemShut {NoStop}%
\bibitem [{\citenamefont {Kingma}\ and\ \citenamefont
  {Welling}(2013)}]{kingma2013auto}%
  \BibitemOpen
  \bibfield  {author} {\bibinfo {author} {\bibfnamefont {D.~P.}\ \bibnamefont
  {Kingma}}\ and\ \bibinfo {author} {\bibfnamefont {M.}~\bibnamefont
  {Welling}},\ }\href@noop {} {\bibfield  {journal} {\bibinfo  {journal} {arXiv
  preprint arXiv:1312.6114}\ } (\bibinfo {year} {2013})}\BibitemShut {NoStop}%
\bibitem [{\citenamefont {LeCun}\ \emph {et~al.}(2015)\citenamefont {LeCun},
  \citenamefont {Bengio},\ and\ \citenamefont {Hinton}}]{lecun2015deep}%
  \BibitemOpen
  \bibfield  {author} {\bibinfo {author} {\bibfnamefont {Y.}~\bibnamefont
  {LeCun}}, \bibinfo {author} {\bibfnamefont {Y.}~\bibnamefont {Bengio}}, \
  and\ \bibinfo {author} {\bibfnamefont {G.}~\bibnamefont {Hinton}},\
  }\href@noop {} {\bibfield  {journal} {\bibinfo  {journal} {Nature}\ }\textbf
  {\bibinfo {volume} {521}},\ \bibinfo {pages} {436} (\bibinfo {year}
  {2015})}\BibitemShut {NoStop}%
\bibitem [{\citenamefont {Li}\ \emph {et~al.}(2020)\citenamefont {Li},
  \citenamefont {Mehta}, \citenamefont {Qian},\ and\ \citenamefont
  {Sun}}]{li2020rethink}%
  \BibitemOpen
  \bibfield  {author} {\bibinfo {author} {\bibfnamefont {T.}~\bibnamefont
  {Li}}, \bibinfo {author} {\bibfnamefont {R.}~\bibnamefont {Mehta}}, \bibinfo
  {author} {\bibfnamefont {Z.}~\bibnamefont {Qian}}, \ and\ \bibinfo {author}
  {\bibfnamefont {J.}~\bibnamefont {Sun}},\ }in\ \href@noop {} {\emph {\bibinfo
  {booktitle} {ICML Workshop on Uncertainty and Robustness in Deep Learning}}}\
  (\bibinfo {year} {2020})\BibitemShut {NoStop}%
\bibitem [{\citenamefont {Baldi}\ and\ \citenamefont
  {Hornik}(1989)}]{baldi1989neural}%
  \BibitemOpen
  \bibfield  {author} {\bibinfo {author} {\bibfnamefont {P.}~\bibnamefont
  {Baldi}}\ and\ \bibinfo {author} {\bibfnamefont {K.}~\bibnamefont {Hornik}},\
  }\href@noop {} {\bibfield  {journal} {\bibinfo  {journal} {Neural Networks}\
  }\textbf {\bibinfo {volume} {2}},\ \bibinfo {pages} {53} (\bibinfo {year}
  {1989})}\BibitemShut {NoStop}%
\bibitem [{\citenamefont {Plaut}(2018)}]{plaut2018principal}%
  \BibitemOpen
  \bibfield  {author} {\bibinfo {author} {\bibfnamefont {E.}~\bibnamefont
  {Plaut}},\ }\href@noop {} {\bibfield  {journal} {\bibinfo  {journal} {arXiv
  preprint arXiv:1804.10253}\ } (\bibinfo {year} {2018})}\BibitemShut {NoStop}%
\bibitem [{\citenamefont {Van Der~Meersch}\ \emph {et~al.}(2023)\citenamefont
  {Van Der~Meersch}, \citenamefont {Deleu},\ and\ \citenamefont
  {Demeester}}]{van2023training}%
  \BibitemOpen
  \bibfield  {author} {\bibinfo {author} {\bibfnamefont {T.}~\bibnamefont {Van
  Der~Meersch}}, \bibinfo {author} {\bibfnamefont {J.}~\bibnamefont {Deleu}}, \
  and\ \bibinfo {author} {\bibfnamefont {T.}~\bibnamefont {Demeester}},\ }in\
  \href@noop {} {\emph {\bibinfo {booktitle} {Associative Memory \& Hopfield
  Networks in 2023}}}\ (\bibinfo {year} {2023})\BibitemShut {NoStop}%
\bibitem [{\citenamefont {Scellier}\ and\ \citenamefont
  {Bengio}(2017)}]{scellier2017equilibrium}%
  \BibitemOpen
  \bibfield  {author} {\bibinfo {author} {\bibfnamefont {B.}~\bibnamefont
  {Scellier}}\ and\ \bibinfo {author} {\bibfnamefont {Y.}~\bibnamefont
  {Bengio}},\ }\href@noop {} {\bibfield  {journal} {\bibinfo  {journal}
  {Frontiers in Computational Neuroscience}\ }\textbf {\bibinfo {volume}
  {11}},\ \bibinfo {pages} {24} (\bibinfo {year} {2017})}\BibitemShut {NoStop}%
\bibitem [{\citenamefont {Laborieux}\ \emph {et~al.}(2021)\citenamefont
  {Laborieux}, \citenamefont {Ernoult}, \citenamefont {Scellier}, \citenamefont
  {Bengio}, \citenamefont {Grollier},\ and\ \citenamefont
  {Querlioz}}]{laborieux2021scaling}%
  \BibitemOpen
  \bibfield  {author} {\bibinfo {author} {\bibfnamefont {A.}~\bibnamefont
  {Laborieux}}, \bibinfo {author} {\bibfnamefont {M.}~\bibnamefont {Ernoult}},
  \bibinfo {author} {\bibfnamefont {B.}~\bibnamefont {Scellier}}, \bibinfo
  {author} {\bibfnamefont {Y.}~\bibnamefont {Bengio}}, \bibinfo {author}
  {\bibfnamefont {J.}~\bibnamefont {Grollier}}, \ and\ \bibinfo {author}
  {\bibfnamefont {D.}~\bibnamefont {Querlioz}},\ }\href@noop {} {\bibfield
  {journal} {\bibinfo  {journal} {Frontiers in Neuroscience}\ }\textbf
  {\bibinfo {volume} {15}},\ \bibinfo {pages} {633674} (\bibinfo {year}
  {2021})}\BibitemShut {NoStop}%
\bibitem [{\citenamefont {Scellier}(2021)}]{scellier2021deep}%
  \BibitemOpen
  \bibfield  {author} {\bibinfo {author} {\bibfnamefont {B.}~\bibnamefont
  {Scellier}},\ }\href@noop {} {\bibfield  {journal} {\bibinfo  {journal}
  {arXiv preprint arXiv:2103.09985}\ } (\bibinfo {year} {2021})}\BibitemShut
  {NoStop}%
\bibitem [{\citenamefont {Cai}\ \emph {et~al.}(2020)\citenamefont {Cai},
  \citenamefont {Kumar}, \citenamefont {Van~Vaerenbergh}, \citenamefont
  {Sheng}, \citenamefont {Liu}, \citenamefont {Li}, \citenamefont {Liu},
  \citenamefont {Foltin}, \citenamefont {Yu}, \citenamefont {Xia} \emph
  {et~al.}}]{cai2020power}%
  \BibitemOpen
  \bibfield  {author} {\bibinfo {author} {\bibfnamefont {F.}~\bibnamefont
  {Cai}}, \bibinfo {author} {\bibfnamefont {S.}~\bibnamefont {Kumar}}, \bibinfo
  {author} {\bibfnamefont {T.}~\bibnamefont {Van~Vaerenbergh}}, \bibinfo
  {author} {\bibfnamefont {X.}~\bibnamefont {Sheng}}, \bibinfo {author}
  {\bibfnamefont {R.}~\bibnamefont {Liu}}, \bibinfo {author} {\bibfnamefont
  {C.}~\bibnamefont {Li}}, \bibinfo {author} {\bibfnamefont {Z.}~\bibnamefont
  {Liu}}, \bibinfo {author} {\bibfnamefont {M.}~\bibnamefont {Foltin}},
  \bibinfo {author} {\bibfnamefont {S.}~\bibnamefont {Yu}}, \bibinfo {author}
  {\bibfnamefont {Q.}~\bibnamefont {Xia}},  \emph {et~al.},\ }\href@noop {}
  {\bibfield  {journal} {\bibinfo  {journal} {Nature Electronics}\ }\textbf
  {\bibinfo {volume} {3}},\ \bibinfo {pages} {409} (\bibinfo {year}
  {2020})}\BibitemShut {NoStop}%
\bibitem [{\citenamefont {Tait}\ \emph {et~al.}(2017)\citenamefont {Tait},
  \citenamefont {De~Lima}, \citenamefont {Zhou}, \citenamefont {Wu},
  \citenamefont {Nahmias}, \citenamefont {Shastri},\ and\ \citenamefont
  {Prucnal}}]{tait2017neuromorphic}%
  \BibitemOpen
  \bibfield  {author} {\bibinfo {author} {\bibfnamefont {A.~N.}\ \bibnamefont
  {Tait}}, \bibinfo {author} {\bibfnamefont {T.~F.}\ \bibnamefont {De~Lima}},
  \bibinfo {author} {\bibfnamefont {E.}~\bibnamefont {Zhou}}, \bibinfo {author}
  {\bibfnamefont {A.~X.}\ \bibnamefont {Wu}}, \bibinfo {author} {\bibfnamefont
  {M.~A.}\ \bibnamefont {Nahmias}}, \bibinfo {author} {\bibfnamefont {B.~J.}\
  \bibnamefont {Shastri}}, \ and\ \bibinfo {author} {\bibfnamefont {P.~R.}\
  \bibnamefont {Prucnal}},\ }\href@noop {} {\bibfield  {journal} {\bibinfo
  {journal} {Scientific Reports}\ }\textbf {\bibinfo {volume} {7}},\ \bibinfo
  {pages} {7430} (\bibinfo {year} {2017})}\BibitemShut {NoStop}%
\end{thebibliography}%

\end{document}